\documentstyle[psfig]{aa}
\def\ltsima{$\; \buildrel < \over \sim \;$}
\def\simlt{\lower.5ex\hbox{\ltsima}}
\def\gtsima{$\; \buildrel > \over \sim \;$}
\def\simgt{\lower.5ex\hbox{\gtsima}}
\def\gsimeq
{\hbox{\raise0.5ex\hbox{$>\lower1.06ex\hbox{$\kern-1.07em{\sim}$}$}}}
\def\lsimeq
{\hbox{\raise0.5ex\hbox{$<\lower1.06ex\hbox{$\kern-1.07em{\sim}$}$}}}

  \title{ X-ray source overdensities in Chandra distant cluster fields: a new probe to map the cosmic tapestry? }

    \author{N.~Cappelluti\inst{1}
      \and
      M.~Cappi\inst{1}
      \and
      M.~ Dadina\inst{1}
      \and
      G.~Malaguti\inst{1}
     \and
      M.~Branchesi\inst{2,3}
     \and
      V. D'Elia \inst{4}
      \and
      G.G.C.~Palumbo\inst{2}}   
\authorrunning{N. Cappelluti et~al.}
\titlerunning{ X-ray source overdensities  in \emph{Chandra}  distant clusters fields. }
  \institute {{IASF-CNR, Sezione di Bologna, Via Gobetti 101, I-40129
               Bologna, Italy\\
               \email{cappelluti@bo.iasf.cnr.it} 
               \and
   Dipartimento di Astronomia Universit\`a di Bologna,
    via Ranzani 1, I--40127 Bologna, Italy\\
        \and 
      IRA-CNR, Via Gobetti 101, I-40129
               Bologna, Italy\\
            \and
                   INAF-Osservatorio astronomico di Roma, Via di Frascati, 33 I-00040 Monteporzio  Catone (Rome), Italy }}

   \offprints{N. Cappelluti (cappelluti@bo.iasf.cnr.it)}
   \date{Received 06/25/2004 / Accepted 09/08/2004 }

\abstract{
The first systematic study  of the serendipitous  X-ray source density around 
10 high $z$ (0.24$<z<$1.2) clusters has been performed with \emph{Chandra}.
A factor $\sim2$ overdensity has been found in 4 cluster  fields, increasing  to 11 the number of 
known  source overdensities around high $z$ clusters.
The result is  statistically highly significant (at $>$2$\sigma$ per field, 
and $>$5$\sigma$ overall) only on scales 8$\arcmin$$\times$8$\arcmin$ 
and  peculiar to cluster fields. The blank field fluctuations, i.e. 
the most sensitive measurements to date of the X-ray cosmic variance on this 
scale, indicate 1$\sigma$ variations within $\sim$15-25\% around the average value.
The first marginal evidence that these overdensities increase 
with cluster redshift is also presented. 
We speculate that the most likely explanation for what has been observed is that 
the x-ray sources are AGNs which trace filaments connected to the clusters. 
Nevertheless other possible explanations such as
an ``X-ray Butcher-Oemler'' effect are not    ruled out by the present results.
If the association of the overdensities with large-scale structures and their 
positive relation with cluster redshift is confirmed, these studies 
could represent  a new, direct, way of mapping the cosmic web and its 
parameters at high $z$.

 }

\begin{document}
\maketitle

\section{Introduction}

Recent studies have reported an unexpected excess of point-like X-ray 
sources on the outskirts of several distant clusters of galaxies 
(Henry \& Briel 1991, Cappi et al. 2001, Pentericci et al. 2002, Molnar et al. 2002, Johnson et al. 2003). 
The source number densities measured around the clusters have been found to exceed 
the values observed in similarly deep observations of fields not centerd on clusters by a factor $\sim$2.  
This is in contrast with  optical and X-ray studies 
which  suggest that AGNs are somewhat rare in
 cluster environments  (Dressler et al. 1985, 1999;  Branchesi et al. in preparation) both in the nearby Universe and at high $z$.  
At least in one  case (3C295 at $z\sim$0.5), the excess number of sources has been demonstrated to be 
 asymmetric  with respect to the cluster, possibly indicating a filament of the large scale structure 
connected with the cluster (D'Elia et al. 2004). 
These pioneering studies   emphasize the potential of  X-ray surveys for tracing the cosmic web at high $z$. 
The statistical significance and nature of these excesses however is not yet clearly established. The main limitations 
are the still poor statistics available from the X-ray studies, as well as the very few optical identifications 
available to date. 
In order to overcome the statistical problems a systematic and uniform 
 analysis of 
the X-ray source population around 10 distant clusters of galaxies was conducted.
 The study was performed using archival {\it Chandra} ACIS-I  observations which are well suited to obtain
X-ray source detections. Throughout this paper, we assume H$_0$= 70 km s$^{-1}$ Mpc$^{-1}$ and $\Omega$=1.
Unless otherwise stated,  errors are at the 1$\sigma$ level. 

\section{Sample selection}

The cluster observations were selected from the $Chandra$ archive using the following criteria: i) clusters should have $z>$0.2 
so that  a wide region around the cluster could be covered, ii) the cluster should be observed with  the ACIS-I CCD (16$\arcmin$ $\times$ 16$\arcmin$) array with iii) the deepest exposure possible ($\sim$100 ks). In total, 10 clusters were 
selected spanning a range of redshifts from 0.2 to 1.2 and with exposures from 30 ks up to  162 ks (see Table 1).

These  fields were compared with  five blank fields (i.e. fields not centered on clusters, hereinafter called ``reference fields'') of similar exposures (see Table 1), namely the ``standard'' and extended Chandra 
deep field south (CDFS and EXT-CDFS), the Hubble deep field north (HDFN), the Groth-Westphal strip  and the ``Bootes'' field. 
In order to sample similar flux levels the analysis of the CDFS and of the HDFN was limited to observation Ids 582  and 3389, respectively.

\begin{table}[!t]
\caption{Clusters and reference fields}
\begin{tabular}{ l r r l l} 
\hline
\hline
Field Name & Obs. Id & Exp. & $z$ & $N^{*}_{\rm HGal}$\\
 & & (ks) &  &   \\
 & & & & \\
\hline
\hline
\underline{\bf Cluster fields:} & & & & \\
MS 1137+6625& 2228   & 110& 0.78  & 1.20\\
CL J0848+4456& 927   & 126 & 0.58 & 2.65\\
RX J0910+5422 & 2227  & 124& 1.11 & 2.09\\
CL J2302+0844 & 918   &110 & 0.72 & 4.91\\
MS 2053-0449 &1667 & 45 &  0.58 & 4.06\\
Abell 2125 & 2207   & 81& 0.24 & 2.74\\
MS 1621+2640 &546 & 30  & 0.43 & 3.57\\
RDCS1252-29 & 4198 & 162 & 1.20 & 2.20 \\
CL J1113.1-2615 &  915 & 105 & 0.72 &5.47 \\
1E 0657-56& 3184   & 90 & 0.29 & 6.53\\
\hline
\hline
\underline{\bf Reference fields:} & & & &\\
CDFS &582 & 130& &0.80 \\
HDFN &3389 & 125 & &1.60 \\
EXT-CDFS-1&   5015& 163 & &0.80 \\
Groth-West. Field &  4357 & 85& &1.28 \\
Bootes-Field & 3130& 115& &0.50 \\
\hline
\hline
\end{tabular}
*Galactic absorption column density, taken from Stark et al. 1992, in units of $\times10^{20}$ cm$^{-2}$. 

\end{table}
\section{Data Reduction and Analysis} 

Data reduction has been performed using the CIAO 3.0.1 software package  (http://cxc.harvard.edu) and
applying the standard technique as follows: photon level 1 event lists
were filtered so as to include only the standard event grades 0,2,4,6, the hot pixels and columns were removed, 
the CTI (Charge Transfer Inefficiency) and gain corrections were applied (http://hea-www.harvard.edu/$\sim$alexey/acis/tgain), flickering pixels were removed and
the intervals with background rates larger than 3$\sigma$ over the quiescent value were excluded. 

The analysis was performed in two energy bands 0.5-2 keV (soft) and 2-10 keV (hard).
The source detection was carried out  chip-by-chip on the unbinned data running the wavelet algorithm in CIAO ({\it wavdetect}, Freeman et al. 2002) 
on a scale of $(\sqrt2)^{i}$ pixel ($i=1,..,8$), i.e. 0.5$\arcsec--$8$\arcsec$. A threshold probability of spurious detections of 10$^{-6}$ has been used (i.e. $\le$1 spurious source per chip) and  those  sources with a detection significance 
$<$3$\sigma$ (see Johnson et al. 2003, and references therein) were excluded.
With this choice the faintest detectable sources have at least $\sim$ 8--9 net counts, which corresponds to a minimum SNR of $\sim$3.
Following the indications given in Manners  (2002), we estimated that this allows uncertainties 
due to the Eddington bias (Eddington 1940) to be lower than $\sim$5$\%$. 
The source net counts were corrected for the vignetting using an exposure map. 
In order to maximize the signal-to-noise of the sources detected in the hard energy band, the analysis was first restricted 
to the 2-7 keV band and then count-rates were extrapolated to the 2-10 keV fluxes (e.g., Giacconi et al. 2001).
The count-rate to flux conversion factor was obtained by assuming a power--law model with 
galactic absorption and photon index ($\Gamma$) estimated from the stacked spectrum of the detected sources in each field.
The typical spectral index found is $\Gamma \sim 1.3-1.4$ in the soft band and $\Gamma \sim 1.6-1.7$ in the hard band.  
Detailed results on this procedure which gives a good estimate of the ``true'' average spectral shape of the detected sources 
is beyond the scope of this work and will be presented in a forthcoming paper.

\section{X-ray source number counts}

\subsection{Sky coverage}
\begin{figure} [!b]
\psfig{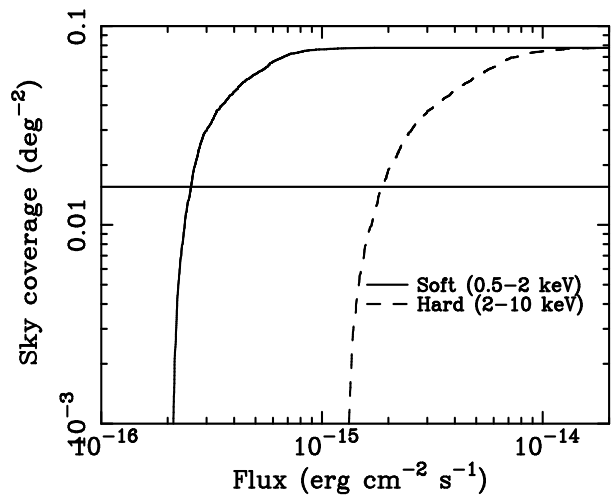}
\caption{Sky coverage as function of flux  in the MS1137+6625 field. The horizontal line represents the lower limit (20\% 
of the total) of the sky coverage which gives the flux limit used in the present work.}
\end{figure}
\begin{figure*}
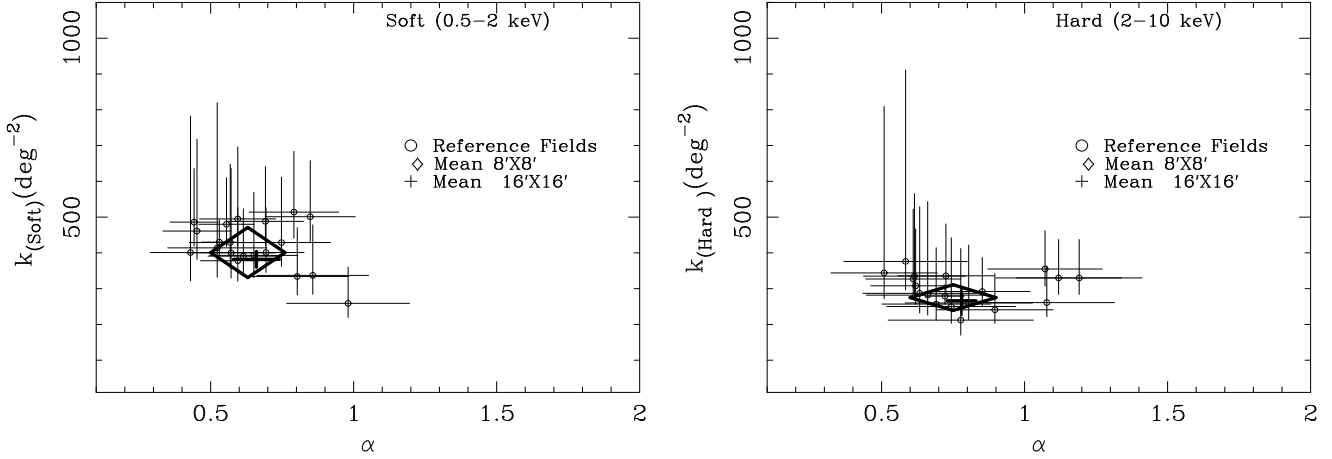

\vspace{1cm}
\begin{tabular}{c c}
\psfig{file=1534f2a.ps,width=8.5cm,height=6cm,angle=-90} &
\psfig{file=1534f2b.ps,width=8.5cm,height=6cm,angle=-90} \\
\end{tabular}
\caption{ Reference fields logN-logS normalization vs. Slope relation in the soft ($Left~Panel$) and  hard ($Right~Panel$) band. Circles indicate best-fit values obtained chip-by-chip (8$\arcmin \times$8$\arcmin$), the thick diamond indicates the mean value error box calculated on scales of 8$\arcmin \times$8$\arcmin$  and cross indicates the mean calculated on  16$\arcmin \times$16$\arcmin$ .  }
\vspace{1cm}
\end{figure*}
Due to instrumental effects, such as vignetting and PSF (Point Spread Function) off-axis degradation, the sensitivity of the ACIS-I detectors varies 
significantly across the field of view. Therefore, to  properly calculate the logN-logS of the flux-limited samples
under investigation, the sky coverage $\Omega$ must be carefully estimated. $\Omega$ defines the area of the detector, 
projected onto the sky, sensitive to a limiting flux $S_{\rm{lim}}$ and was calculated as follows.
 For each chip a background map was constructed from the image 
after excluding the detected sources and filling the remaining 
``holes''  (using  the {\it dmfilth} tool in CIAO) with the surrounding average background. The maps were then re-binned into 64$\times$64 matrices of $\sim$30$\arcsec$ per pixel
to smooth the local background variations and   to sample the larger size of the off-axis PSF.
Following Johnson et al. (2003), the background level in each bin was used to estimate the number of counts needed to obtain 
a $>$3$\sigma$ detection (with $wavdetect$). The flux limit $S_{\rm{lim}}$ (in units of  erg cm$^{-2}$ s$^{-1}$) per image bin was then calculated using the formula:

\begin{equation}
S_{\rm{lim}}=3\times B \times V \times f \times (1+\sqrt{0.75+(B \times D^{2})})/t    
\end{equation}
where $B$ are the background counts, $V$ is the vignetting correction factor (given by the exposure map),
$f$ is the count-rate to flux conversion factor, $D$ is the estimated source diameter and $t$ is the exposure time.
The sky coverage at a given flux $S_{\rm{lim}}$ is then given by the sum of all the regions of the detector with a flux limit larger 
than $S_{\rm{lim}}$ and then  converted to deg$^{2}$.
As shown in Fig. 1, the sky coverage rapidly decreases near the survey limiting flux for both the soft (continuous line) and 
hard (dashed line) energy bands. In order to prevent incompleteness effects  only   sources 
with  flux greater than  the flux  corresponding to  20\% of the sky coverage (see horizontal line in Fig. 1) were  included in the analysis.

\subsection{logN-logS}

The cumulative source number counts $N(>S)$ have been calculated chip-by-chip following the basic method described in 
Gioia et al. (1990) and using the following formula:  

\begin{equation}
N(>S)=\sum_{i=1}^{n}\frac{1}{\Omega_{i}} ~~  \rm{deg}^{-2}
\end{equation}
where $n$ is the number of detected sources  and $\Omega_{i}$ is the sky
coverage corresponding to the flux of the $i$-th source.

A maximum likelihood method as given in Crawford et al. (1970) and Murdoch et al. (1973)  was then used to fit the unbinned data with a single 
power law model of the form:

\begin{equation}
N(>S)=k(\frac{S}{S_{0}})^{-\alpha} ~~\rm{deg}^{-2}
\end{equation}
where $k$ and $\alpha$ are the normalization and slope of the power law.

The only free parameter of the fit is $\alpha$, while the normalization $k$ is computed assuming 
that the predicted source  counts at the flux limit  are  equal to what is observed. 
Errors in $k$ were then calculated using the upper and lower limits of $\alpha$. 
The value of $k$ was calculated at flux levels of S$_{0}$=2$\times10^{-15}$ erg cm$^{-2}$ s$^{-1}$ in the soft band and 
 S$_{0}$=1$\times10^{-14}$ erg cm$^{-2}$ s$^{-1}$ in the hard band; for the exposures considered 
here these correspond  to the centers of the sampled flux intervals.

\section{Results}

The logN-logS of sources around the clusters has been calculated chip-by-chip, as well as over the whole 
(4 chips) FOV, and  compared to the logN-logS of the reference fields.

\subsection{Reference  Fields}

As detailed above (see Sect. 2 and Table 1), 5 blank fields  
(20 chips, i.e., a sky coverage of $\sim0.4$ deg$^{2}$) have been used as reference fields.
Best-fit results of the logN-logS obtained chip-by-chip (8$\arcmin$ $\times$ 8$\arcmin$) are reported in Fig. 2.
As shown in Fig. 2, the  best-fit normalizations $k$  are centered around the error-weighted means k$_{\rm{soft}}$=401$^{+31}_{-13}$ deg$^{-2}$ 
 (standard deviation $\sigma_{\rm{soft}}$=70$\pm$3 deg$^{-2}$) and  k$_{\rm{hard}}$=275$^{+22}_{-9}$ deg$^{-2}$  ($\sigma_{\rm{hard}}$=36$\pm$5 deg$^{-2}$), in
the soft and hard energy bands, respectively.
These average values (marked with diamonds in Fig. 2) are in very good agreement with 
other recent studies (e.g. Moretti et al. 2003, Stern et al. 2002, 
Brandt et al. 2001, Giacconi et. al 2001, Rosati et al. 2002, Tozzi et al. 2001).

Field to field variations, otherwise known as ``cosmic variance'' 
(see e.g. Soltan et al. 2001, Vikhlinin et al 1995), are  within 
a  spread of $\sim$20-25\% around these  mean values. This 
 is in agreement 
with the few other recent studies performed at similar fluxes (e.g. Yang et al. 2003, Wang et al. 2004). 
Slope indeces are always consistent with the mean values of 0.63$\pm{0.13}$ 
in the soft and 0.75$\pm{0.15}$ in the hard energy bands.  

\begin{figure} [!t]
\psfig{file=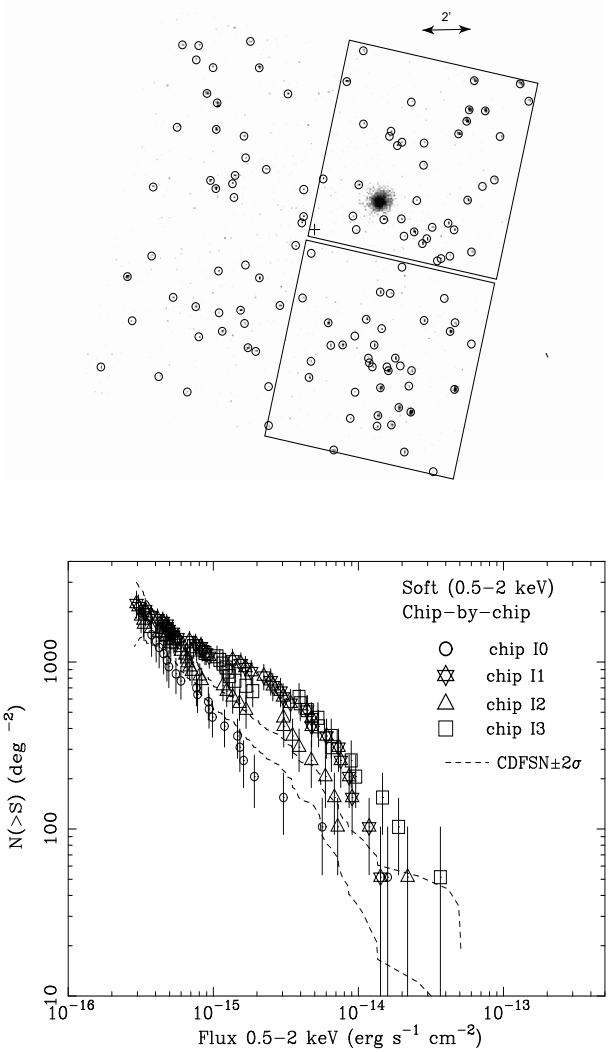,width=8.cm,height=6.5cm,angle=0}
\vspace{1cm}
\psfig{file=1534f3b.ps,width=8.cm,height=6.5cm,angle=270}
\caption{($Top$) MS1137+6625 0.5-2 keV image, circles represent the detected sources, the cross indicates the aim-point. ($Bottom$) 0.5-2 keV chip-by-chip logN-logS of the MS1137+6625 field. Dashed line represents the 2$\sigma$ confidence level of the CDFS+HDFN logN-logS. }
\end{figure}
\begin{table}[!b]
\begin{center}
\caption{Statistical properties of the overdensities}
\begin{tabular}{l l l l l l}
& & Soft &  & Hard \\
\hline
\hline 
Cluster & Chip & N$^{\circ}$ & P &  N$^{\circ}$ & P\\
    &  &   of $\sigma$& &    of $\sigma$ & \\   

&& (1)& (2)&(3)&(4)\\
\hline
\hline
MS1137+6625 & I1 &2.47  &1.35\% & 3.10 & 0.20\%\\
             & I3$^{\dag}$ &2.36 &1.83\% & 2.57 &1.02\%\\
CLJ0848+4456 &I2 &2.25& 2.44\% & 2.41& 1.60\%\\
RDCS0910+5422& I3$^{\dag}$ & 2.20&2.78\% &2.97&0.30\%\\
             & I2 & 2.03& 4.24\% & *&*\\ 
MS2053-0449 & I0&* &* & 2.66& 0.78\%\\
      & I3$^{\dag}$& 2.23&2.57\% &2.46&1.43\%\\
           
\hline
\hline
\end{tabular}
\end{center}
\begin{footnotesize}
(1) Significance of the excess in units of $\sigma$ (Soft).\\(2) Probability of upward Poisson fluctuation of reference field distribution (Soft).\\(3) Significance of the excess in units of $\sigma$ (Hard)\\(4) Probability of upward Poisson fluctuation of reference field distribution(Hard).\\
$^{\dag}$ Chip containing the cluster.\\
\end{footnotesize}
\end{table}

\subsection{Cluster fields}
\begin{figure*}
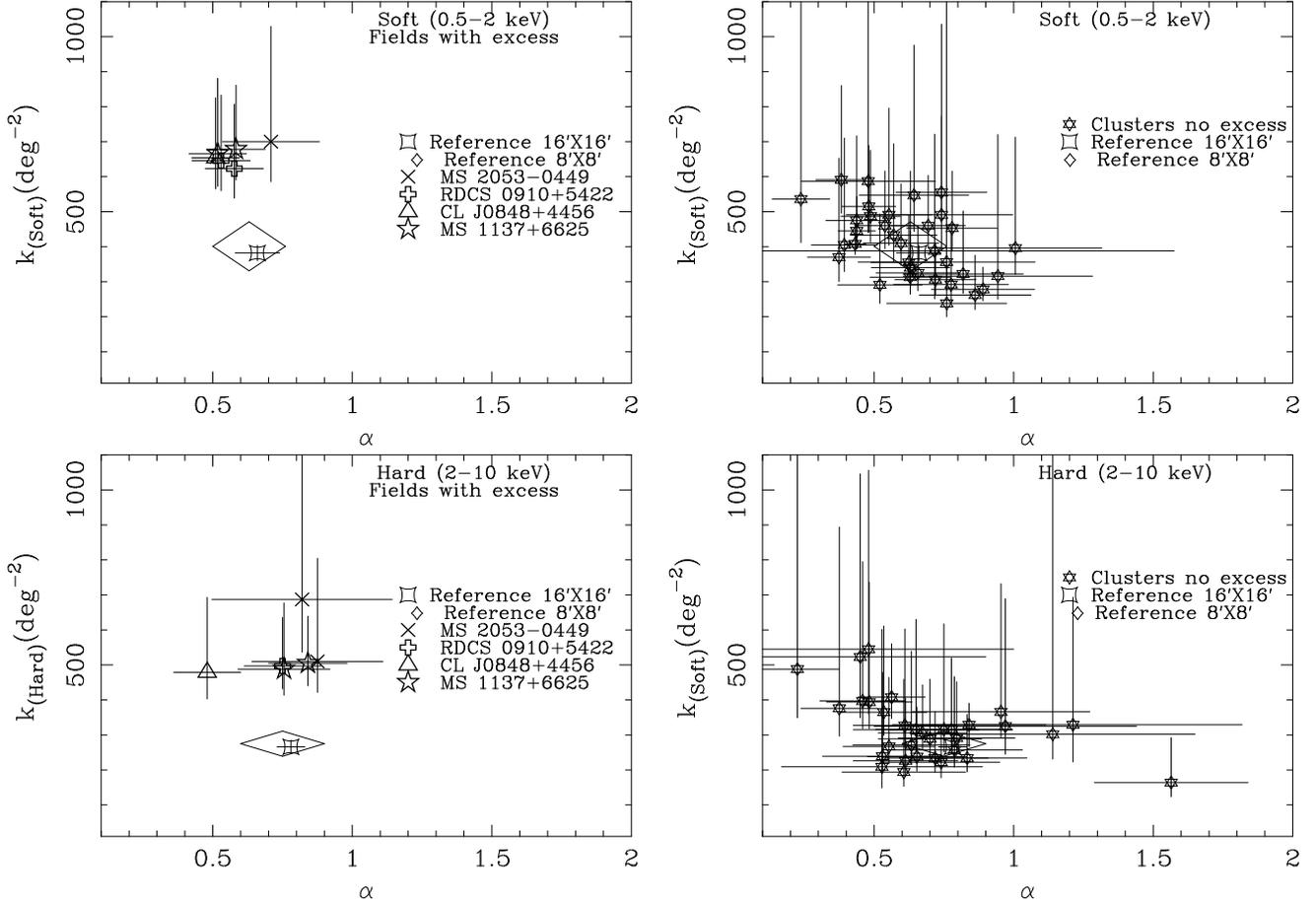

 \vspace{2cm}
\begin{tabular}{c c}
\psfig{file=1534f4a.ps,width=8.5cm,height=6cm,angle=-90} & 
\psfig{file=1534f4b.ps,width=8.5cm,height=6cm,angle=-90} \\
\psfig{file=1534f4c.ps,width=8.5cm,height=6cm,angle=-90} &
\psfig{file=1534f4d.ps,width=8.5cm,height=6cm,angle=-90} \\
\end{tabular}
\caption{ logN-logS normalization vs. slope relations for overdense cluster fields in the soft ($Top~left~panel$) and   hard band ($Bottom~left~panel$). The same relation in cluster fields with non-significant or no excesses in the soft ($Top~right~panel$) and   hard band ($Bottom~right~panel$).  }

\end{figure*}
\begin{figure*} 
\begin{tabular}{c c}
\psfig{file=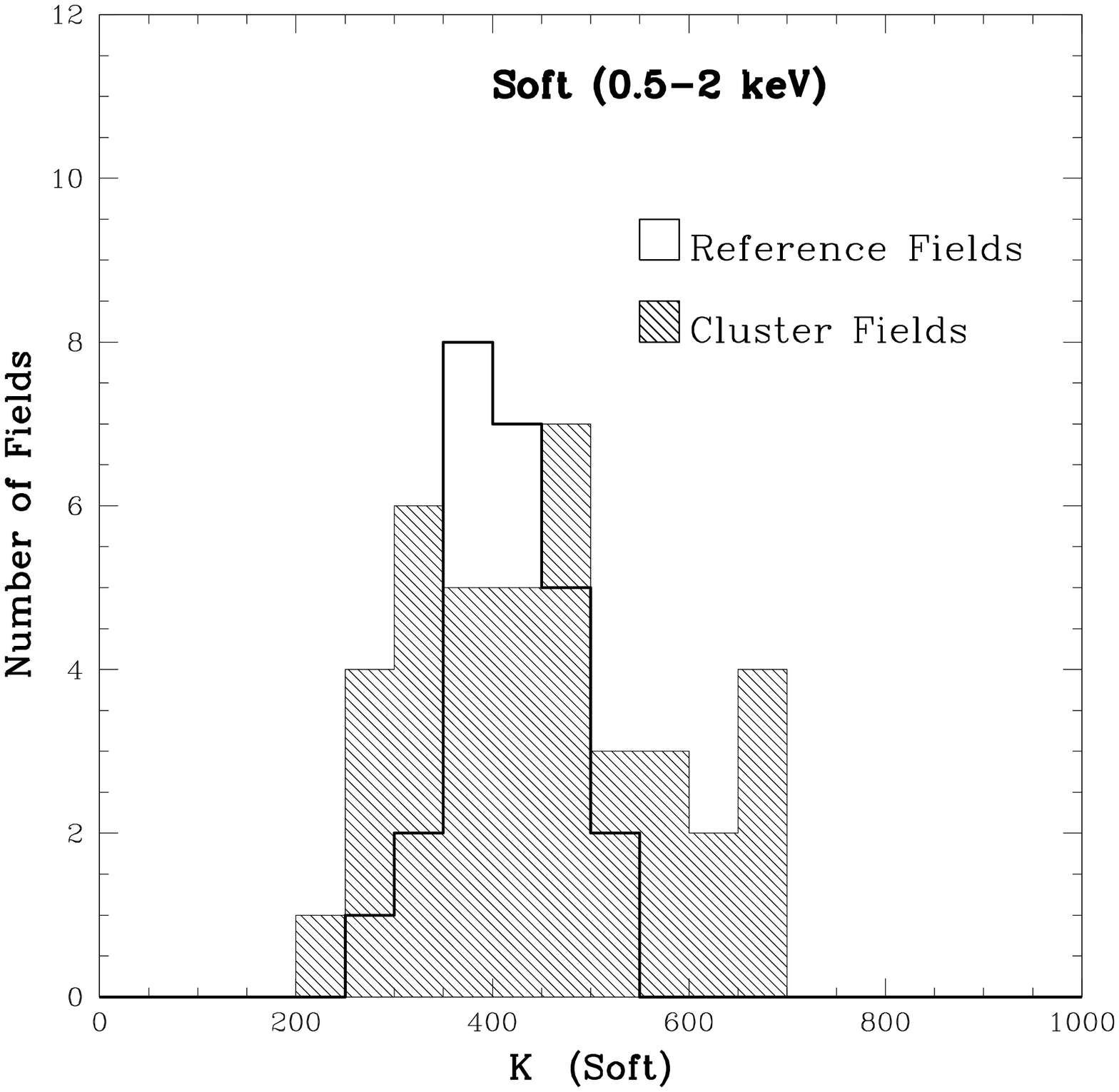,width=8.5cm,height=6cm,angle=0} & 
\psfig{file=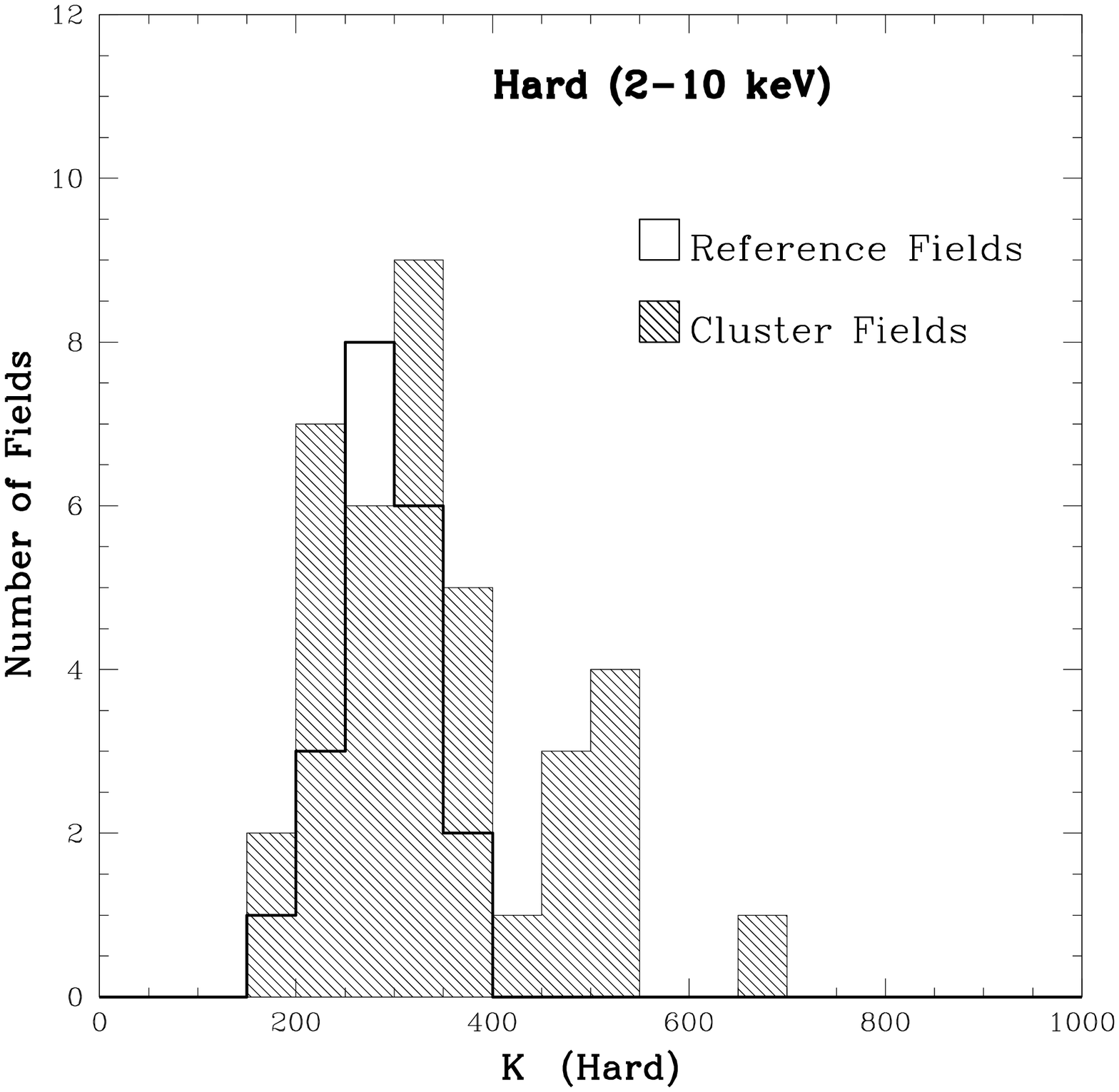,width=8.5cm,height=6cm,angle=0} \\
\end{tabular}
\caption{ Distribution of the logN-logS normalizations in cluster and reference fields, 
in the soft ($Left~Panel$) and hard ($Right~Panel$) 
energy bands.}
\end{figure*}

As an example we first show the results obtained in the field of MS 1137+6625.
Fig. 3 ($Top$) clearly shows that the source density in the two boxed chips (one of which contains MS1137+6625) is 
about twice  that in the other two chips. Chip-by-chip logN-logS have
been calculated and are shown in Fig. 3. They are 
compared to the CDFS+HDFN logN-logS (CDFSN, $\pm2\sigma$ deviations) that has a best fit set of parameters  within 5\% of the  mean value given in Sect. 5.1. 
 For two chips the  logN-logS show a source surface density twice  that in the CDFSN.   

As detailed above (see Sect. 2 and Table 1), a similar analysis was performed on 10 cluster fields  for a total of  40 chips,
 i.e. a sky coverage of $\sim0.8$ deg$^{2}$.
Best-fit normalizations and slopes for the logN-logS in the soft and hard energy bands are shown in Fig. 4, and compared to the 
average values of the reference fields (marked with diamonds in the figures).
In six chip (from 4 cluster fields)  excesses of a factor $\sim$1.7-2 with a significance $>2\sigma$ are
detected. 
These are clearly seen in Fig. 4 ($Left~Panels$), and cluster-by-cluster statistics are given in Table 2.
The excesses are located in the chip containing  the cluster and in the adjoining chip. In the Lynx field (here named CL J0848+4456) the  excess is observed in the only  chip without a cluster (i.e. in this field three clusters were detected).

The other 6 fields  (24 chips) do not show any significant ($>2\sigma$) excess. 
Nevertheless, in four  out of these six  cluster fields, given the low statistics, factor $\sim2$ 
deviations cannot be statistically  ruled out either.
Altogether, we find {\it at least} 4 clusters out of 10 with significant overdensities on angular scales of 
$8\arcmin \times 8\arcmin$ or $8\arcmin \times 16\arcmin$.

It is worth noting  that when calculated over the entire FOVs, these excesses are  
smoothed out and the logN-logS become statistically consistent with the reference fields on the 
same angular scales.

Fig. 5 shows the  histograms of the normalization $k$ in  cluster and reference fields (chip-by-chip).
The histograms clearly show that the distributions have an almost Gaussian shape in the reference fields, both 
in the soft (Fig. 5, $Left$)  and hard (Fig. 5, $Right$)  energy bands. Instead, in the cluster fields, the distributions 
have a prominent tail at high values of k, containing about 20-25\% of the fields. 
The significance of the tail can be quantified with a KS test which yields a probability that the 
two distributions are extracted from the same  parent population of 9.03$\times10^{-3}$ in the soft band 
and 2.55$\times10^{-6}$ in the hard band.

\subsection{Overdensities as a function of z}

A very  interesting result obtained from the present analysis is the apparent correlation found 
between the  amplitude of the overdensities  and the cluster redshifts, as shown in Fig. 6. 
In this figure, the  the (2-10 keV) $k$  normalized to the value of the reference 
fields increases with the cluster redshift. The correlation is significant at a level of $\sim96\%$  
yielding a correlation parameter r=0.69 for 10 independent points. The best-fit linear relation is here 
R=(0.63$\pm$0.24)$z$+0.9$\pm$0.2 in the hard band while  a similar relation 
R=(0.55$\pm0.10$)$z$+1.06$\pm0.16$ is obtained in the soft band.
 
\begin{figure}[!t]
\psfig{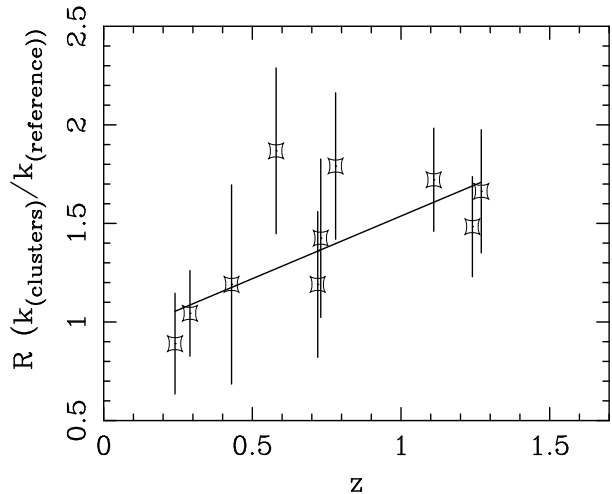}
\caption{Overdensities ($k_{\rm{clusters}}/k_{\rm{reference}}$) as a function of cluster redshift, 
in the hard band.}
\end{figure} 

\section{Discussion}

The surface density of the X-ray serendipitous sources around 10 high $z$ clusters has been investigated with 
$Chandra$ and compared to the surface density of 5 reference cluster-free fields   observed with similar exposure times.
Four of the cluster fields show a factor $\sim$2 overdensity of sources at a $>$2$\sigma$ significance level, while there clearly are no similar deviations present in the reference fields.

The fact that, as shown in Fig. 5, a fraction of $\sim$25\% of all fields near clusters shows overdensities 
demonstrates that this effect is unlikely to be associated  to random statistical fluctuations constitutes a peculiar property of  cluster fields. 
With the present work, the number of known high $z$ clusters showing such 
excesses increases to eleven\footnote{The other 7 known cases being A2256 (Henry \& Briel  1991), 
3C295 and RXJ0030+2618 (Cappi et al. 2001), MRC1138-262 (Pentericci et al. ,2002), 
A2104 (Martini et al., 2002) MS 1054-0321 (Johnson et al., 2003) and A1995 (Molnar et al. 2002).}. 
At a first glance, this contradicts what is reported in Kim et. al (2004) who 
find no significant difference between cluster and cluster-free fields (62 fields in total).
Those authors, however, searched for source count differences on larger angular scales\footnote{Indeed, Kim et al. (2004) 
analyzed a total of 29 fields with clusters and 33 without and calculated the logN-logS 
either over a circular area per single FOV of radius 6.6 arcmin (i.e. with a sky coverage 
$\sim$2 times larger than in this work), or adding together all fields reaching a cumulative
sky coverage of $\sim$1 deg$^2$.} than in the present work. This may have had the effect of 
smoothing any excess signal. 
In support of the above argument we find that  integrating our data over the whole (16$\arcmin$$\times$16$\arcmin$) 
ACIS-I FOVs, we  find no statistically significant difference between  fields either. 
In other words, this is consistent with an increase, in cluster fields, of the angular correlation signal 
toward lower angular scales (see e.g. D'Elia et al. 2004). 

It should also be mentioned here that the present study yields one of the most sensitive measurements of 
the X-ray cosmic variance in empty fields available to date. Indeed,  on scales of 
8$\arcmin$$\times$8$\arcmin$ as reported in $Sect. 5.1$ and clearly shown
in Fig. 5, we find (1$\sigma$) fluctuations in number counts of the order of 17\% and 13\% in the soft 
and hard band, respectively. On scales 16$\arcmin$$\times$16$\arcmin$, 
these values decrease to   8\% and 7\% in the soft and hard band, respectively. 

Considering that, if the detected X-ray sources are associated with the clusters themselves, their average 
luminosity must be of the order of $\sim$10$^{42-44}$ erg/s at $z$=0.5 and 
$\sim$10$^{43-44.5}$ erg/s at $z$=1.2, they are most likely to be active galactic nuclei.
This is also supported by the power--law shape of their summed spectra (see Sect. 3).
Moreover, if connected with the clusters, the angular scales on which the 
overdensities emerge correspond to linear distances of the order of 3-7 Mpc (at $z$=$0.5$ and $1.2$, 
respectively). This is somewhat greater than the typical dynamical radius of $<$3 Mpc for a 
massive cluster. 

A number of explanations have been proposed in the literature (e.g. Cappi et al. 2001) 
to explain this kind of  excess of X-ray sources, e.g.: gravitational lensing, triggering of 
AGN/starburst activity or filamentary clustering.

The hypothesis that some of the detected sources are magnified by gravitational 
lensing  seems rather unlikely because i) the shapes of the logN-logS do not 
get steeper at fainter fluxes, as  required by lensing in order to obtain a significant increase in the 
surface density of faint sources, and ii) the most distant  clusters have $Einstein$ angles of the 
order $<1\arcmin$, much smaller than the angular  scale of the excess. This further confirms previous findings 
by Cappi et al. (2001) and Johnson et al. (2003), based on the calculations done by 
Refregier \& Loeb (1997).

Butcher \& Oemler (1984) discovered 
that the fraction of blue (active) galaxies is very low ($<<$10\%) in nearby clusters but
increases  linearly with the cluster redshift (percentages of several tens at $z$=0.5).
This may well explain, at least qualitatively,  the redshift-dependence found here in the 
X-ray band (Fig. 6).
Simulations suggest that this effect could be caused by the interaction 
of the intra-cluster  with the interstellar medium in those galaxies which 
are falling  into the cluster potential well (see e.g Evrard et al. (1991)), 
thereby triggering  AGN/Starburst activity. However, such an explanation faces the problem  that 
the scales involved here ($>$3 Mpc) are substantially greater than typical virial radii, 
thus it is not clear how much diffuse gas would be available at those distances to 
trigger such an interaction. 

The third hypothesis, that the excesses could represent filaments of the large-scale structure of 
the Universe, is in our opinion the most likely explanation. 
In the current vision of the Universe, clusters lie in the knots of 
a filamentary web-like network of galaxies and Dark Matter 
(see e.g. Peebles, 1993). This has been demonstrated by both hydrodynamic simulations 
(Bond et al. 1996,Colberg et al. 2000) and by extensive optical imaging and spectroscopic observations of galaxies 
around clusters (e.g. Gioia et al. 1999, Ebeling et al. 2004, Le F{\` e}vre et al. 2004).

If AGNs trace galaxies (as observations clearly indicate, see for example Fig. 13 in Kauffmann et al, 2004),
 it is thus possible that we are observing AGNs in filaments 
connected with the clusters themselves.
The detailed study by D'Elia et al. (2004) of the field surrounding the cluster 3C295 
at $z$=0.5 clearly shows a strong and asymmetric clustering of X-ray sources on scales 
of a few arcmin. This dataset is, to our opinion, one of the best pieces of evidence to date for a 
filament at $z\sim$0.5 traced by X-ray emitting AGNs. Spectroscopic identification of the sources is being 
carried out and some redshifts close to that of the cluster give a first 
hint of a spatial connection between the overdensity and the cluster 
itself (D'Elia, private communication). 
Ohta et al. (2003) recently discovered in the optical band an overdensity of bright QSOs 
(a QSO-supercluster) and several groups of galaxies in the cluster 
field CL J0848+4456 at z$\sim1.2$. This cluster field also shows 
an excess of X-ray sources (Fig. 4), and thus could be related to the large-scale 
structure at $z\sim$1.2. Unfortunately, these optically identified quasars  are  spread 
over an area (30$\arcmin$$\times$30$\arcmin$) much larger 
than the Chandra FOV (see Fig. 3 in Ohta et al. 2003). Therefore only four 
identifications (one QSO at $z\sim$1.26, one at $z\sim$0.88 and two at $z\sim$0.58) are in 
common making  conclusions uncertain.  Assuming  that the excesses are indeed linked to the large scale structure of the Universe, the  redshift-dependence of the overdensities (Fig. 6) 
could, in turn, be explained  as
a geometrical effect.

In order to  firmly establish 
whether the X-ray overdensities found here are associated to large-scale structures 
at high $z$, extensive spectroscopic follow-up observations\footnote{Redshift 
 measurements are necessary to obtain 3-D information to firmly confirm/exclude the large-scale structure hypothesis.
Such investigations are being planned but,  given 
the optical faintness of most X-ray sources, will require observing time on 8 meter-class 
telescopes.} similar to those performed by 
e.g. Gioia et al. (1999), Ebeling et al. (2004), are required.

But clearly  X-ray studies of  source overdensities  around high $z$ clusters 
might be a rather efficient way of mapping the cosmic web at high $z$.

\section{Conclusions}
The first systematic study  of the serendipitous  X-ray source density around 
high $z$ galaxy clusters has been performed with \emph{Chandra}. For the first time the results of  study have been compared to
a sample of cluster-free fields which also allowed  a detailed measurement of the cosmic variance.
Forty percent of the cluster fields present strong statistically significant overdensities. 
The first, although not robust, evidence for a  redshift dependence of these excesses has also been presented.  

 We speculate that filaments connected to the clusters are the most 
likely astrophysical interpretation of the above results. However, thought less likely, 
other effects such as an ``X-ray Butcher Oemler effect'' cannot be completely ruled out.

\section{Acknowledgments}

We wish to thank P. Mazzotta, A. Vikhlinin, I. Gioia and  F. Fiore
 for many   
useful comments. N.C thanks R. Della Ceca who provided  the logN-logS fitting package,  Matteo Genghini, Enrico Franceschi  and Fulvio Gianotti for software--related assistance.
The entire $Chandra$ team, who made these observations possible, is acknowledged. 
Financial support from ASI and MIUR is kindly acknowledged.

{}

\end{document}